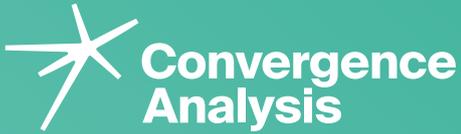

# Training Data Attribution (TDA): Examining Its Adoption & Use Cases

BY DERIC CHENG, JUHAN BAE, JUSTIN BULLOCK, DAVID KRISTOFFERSON

JULY 2024

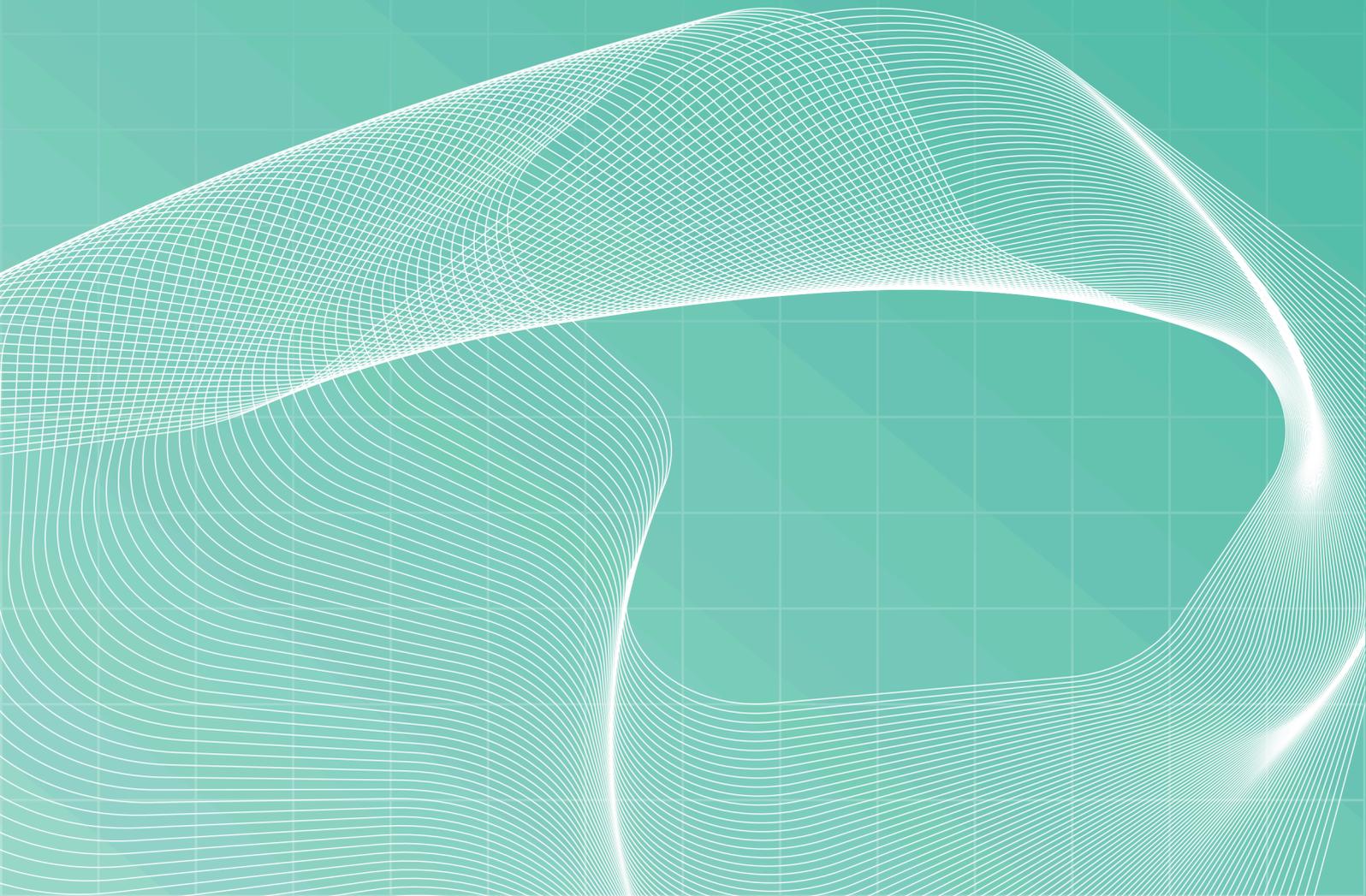

# Table of Contents

NOTE

This report was conducted in June 2024 and is based on research originally commissioned by the Future of Life Foundation (FLF). The views and opinions expressed in this document are those of the authors and do not represent the positions of FLF.

# Executive Summary

This report investigates *Training Data Attribution (TDA)* and its potential importance to and tractability for reducing extreme risks from AI. TDA techniques aim to identify training data points that are especially influential on the behavior of specific model outputs. They are motivated by the question: how would the model's behavior change if one or more data points were removed from or added to the training dataset?

Report structure:

- First, we discuss the plausibility and amount of effort it would take to bring existing TDA research efforts from their current state, to an efficient and accurate tool for TDA inference that can be run on frontier-scale LLMs. Next, we discuss the numerous research benefits AI labs will expect to see from using such TDA tooling.

- Then, we discuss a key outstanding bottleneck that would limit such TDA tooling from being accessible publicly: AI labs' willingness to disclose their training data. We suggest ways AI labs may work around these limitations, and discuss the willingness of governments to mandate such access.

- Assuming that AI labs willingly provide access to TDA inference, we then discuss what high-level societal benefits you might see. We list and discuss a series of policies and systems that may be enabled by TDA. Finally, we present an evaluation of TDA's potential impact on mitigating large-scale risks from AI systems.

Key takeaways from our report:

- Modern TDA techniques can be categorized into three main groups: retraining-based, representation-based (or input-similarity-based), and gradient-based. Recent research has found that gradient-based methods (using influence functions) are the most likely path to practical TDA.

- The most efficient approach to conduct TDA using influence functions today has training costs on par with pre-training an LLM. It has significantly higher (but feasible) storage costs than an LLM model, and somewhat higher per-inference costs.

- Based on these estimates, TDA appears to be no longer infeasible to run on frontier LLMs with enterprise-levels of compute and storage. However, these techniques have not been tested on larger models, and the accuracy of these optimized TDA techniques on large models is unclear.

- Compressed-gradient TDA is already plausible to be used on fine-tuned models, which have orders of magnitude fewer training examples and parameters (on the order of millions or billions rather than hundreds of

- billions).
- Timing to achieve efficient and accurate TDA on frontier models is likely between 2-5 years, depending largely on specific incremental research results and amount of funding / researchers allocated to the space.
- Efficient TDA techniques will likely have a substantial positive impact on AI research and LLM development, including the following effects:
    - Mitigating the prevalence of hallucinations and false claims
    - Identifying training data that produces poor results (bias, misinformation, toxicity), improved data filtering / selection
    - Shrinking overall model size / improving efficiency
    - Improved interpretability & alignment
    - Improved model customization and editing
- AI labs are likely already well-incentivized to invest in TDA research efforts because of the benefits to AI research.
- Public access to TDA tooling on frontier AI models is limited primarily by the unwillingness / inability of AI labs to publicly share their training data.
    - AI labs currently have strong incentives to keep their training data private, as publishing such data would have negative outcomes such as:
        - Reduced competitive advantages from data curation
        - Increased exposure to legal liabilities from data collection
        - Violating privacy or proprietary data requirements
    - AI labs may be able to avoid these outcomes by selectively permitting TDA inference on certain training examples, or returning sources rather than the exact training data.
    - Governments are highly unlikely to mandate public access to training data.
- If AI labs willingly provided public access to TDA, you could expect the following benefits, among others:
    - Preventing copyrighted data usage.
    - Improved fact checking / content moderation
    - Impacts on public trust and confidence in LLMs
    - Accelerated research by external parties
    - Increased accountability for AI labs
- AI labs appear largely disincentivized to provide access to TDA inference, as many of the public benefits are disadvantageous for them.
    - Governments are highly unlikely to mandate public access to TDA.
    - It seems plausible that certain AI labs may expose TDA as a feature, but that the majority would prefer to use it privately to improve their models.

- Several systems that could be enabled by efficient TDA include:
  - Providing royalties to data providers / creators
  - Automated response improvement / fact-checking
  - Tooling for improving external audits of training data
  - Content attribution tooling for LLMs, though it is unlikely to replace systems reliant on RAG
- We believe that the most promising benefit of TDA for AI risk mitigation is its potential to improve the technical safety of LLMs via interpretability.
  - There are some societal / systematic benefits from TDA, and these benefits may be a small contributing factor to reducing some sources of risk. We don't think these appear to move the needle significantly to reduce large-scale AI risks.
  - TDA may meaningfully improve AI capabilities research, which might actually increase large-scale risk.
  - TDA may eventually be highly impactful in technical AI safety and alignment efforts. We'd consider TDA's potential impact on technical AI safety to be in a similar category to supporting mechanistic interpretability research.

# What is Training Data Attribution (TDA) for Large Language Models (LLMs)? What are the most promising approaches?

Training data attribution (TDA) techniques are designed to identify the training data points that significantly influence a model's output. They are often motivated by the counterfactual question: how would the model's behavior change if one or more data points were removed from or added to the training dataset? Given a query example, $z_{query}$ (e.g., a test sequence or a prompt & completion pair), the model behavior is typically quantified using the measurement $f(z_{query}, \theta)$, selected based on metrics relevant to the analysis, such as loss, logits, or log probability. For example, if the measurement is set to loss, TDA techniques aim to identify the training example $z_i$ that leads to the most substantial change in the loss on a query sequence when the model is trained without $z_i$ in the training dataset[1].

Modern TDA techniques can be categorized into three main groups: (1) retraining-based, (2) representation-based (or input-similarity-based), and (3) gradient-based. For a comprehensive overview of TDA, we refer readers to Hammoudeh & Lowd (2022) and Mucsanyi et al. (2023).

Retraining-based approaches, such as Shapley-value estimators (Shapley et al., 1953), downsampling (Feldman & Zhang, 2020), Datamodels (Ilyas et al., 2022), Data Banzhaf (Wang & Jia, 2023) approximate the effect of removing a data point (or a group of data points) by repeatedly retraining models. For instance, Datamodels train over 100,000 networks using different subsets of the dataset, building a linear regression model that predicts how the measurement changes as we exclude some portion of the data examples. Although effective, these approaches would be impractical for large language models (LLMs) due to the significant retraining involved. However, they can be valuable in formulating the ground truth for TDA techniques for smaller models or datasets (e.g., a 100 million parameter model on a GLUE fine-tuning dataset).

Representation-based techniques evaluate the relevance between a training and query data point by examining the similarity of their hidden representations (Caruana et al., 1999). The hidden representation can be the final hidden activation or a concatenation of activations from all layers, and the similarity is typically quantified using dot product or cosine metrics (Hanawa et al., 2020). For example, Rajani et al. (2020) performed data attribution using the "[CLS]" representation of the last layer of BERT. It is worth noting that the hidden states need not necessarily be computed using the model of interest. Singla et al. (2023) employed a self-supervised model for data attribution in image classification models, making the attribution process model-agnostic. These representation-based techniques offer computational advantages compared to other attribution techniques, as they only require forward passes through some network. In an extreme case, some works compute the similarity between the data points themselves, using techniques such as TF-IDF (Grosse et al., 2023) or BM25 (Akyürek et al., 2022), which does not involve any neural networks.

---

[1] For a tutorial on recent methods in TDA techniques, see the resource: Data Attribution at Scale.

Representation-based or input-similarity-based methods are effective in finding similar data examples. On image classification tasks, given a query data point, these methods have been shown to be effective in identifying training data points belonging to the same class (Hanawa et al., 2020). However, they lack a clear connection to the counterfactual. Similar training data examples to the query do not necessarily correspond to a significant change in the query's measurement when those particular training data examples are removed[2]. For instance, Singla et al. (2023) qualitatively observed that representation-based methods identify more visually similar training images compared to the retraining-based method (Datamodels), but perform worse in counterfactual evaluations. Park et al. 2023 made similar observations on the mT5 model for the fact-tracing benchmark (Akyürek et al., 2022). Moreover, due to a lack of evaluation benchmarks in the TDA community, several design choices remain unclear: (1) whether to use intermediate layers of Transformers with averaged activations over tokens (Akyürek et al., 2022), (2) whether to use the last layer activation with the representation of a particular token (Rajani et al., 2020), (3) which model to use for extracting embeddings (Singla et al., 2023).

Moreover, the naive implementation of representation similarity for LLMs is computationally impractical. For a new given query $z_{query}$, it necessitates performing a forward pass through all pre-training sequences (this has to be repeated for each new set of queries). One feasible approach is to cache all (potentially projected) hidden states on disk and use an approximate nearest neighbor search (Johnson et al., 2017; Rajani et al., 2020). It is worth noting that there also exist several methods that perform TDA with only forward passes (Ladhak et al., 2023; Ko et al., 2024). However, these methods do not have a straightforward caching mechanism and are likely not feasible at a larger scale, as they require repeated forward passes for each query example.

Gradient-based methods estimate the sensitivity of the final model parameters to individual training data points. Gradient-based methods can further be divided into implicit-differentiation-based, unrolled-differentiation-based, and TracIn-based approaches. The implicit-differentiation-based TDA, most notably influence functions (Hampel, 1974; Koh & Liang, 2017), uses the Implicit Function Theorem (Krantz & Parks, 2002) to estimate the sensitivity of downweighting a data point on the optimal solution. While they are well-studied for linear models, they lack a clear connection to the counterfactual as they make assumptions such as uniqueness of and convergence to the optimal solution. Park et al. 2023 showed some empirical connection to the counterfactual on neural networks, and Bae et al., 2022 theoretically demonstrated the exact object that implicit-differentiation-based methods approximate. It is noteworthy that these approaches are fundamentally inapplicable to analyzing multi-stage procedures such as continual learning or foundation models (Guu et al., 2023). For instance, they cannot answer questions such as "What is the impact of removing this training example from early versus late stages of training?".

---

[2] See Søgaard et al. 2021 for an example in toy binary classification setting.

The other two approaches – unrolled differentiation and TracIn – are not practical for attribution in LLMs. While unrolled differentiation (Hara et al., 2019; Chen et al., 2021) provides a more fundamental formulation to approximate the counterfactual for neural networks, it brings significant computational and memory overhead. Similarly, TracIn-based approaches (Pruthi et al., 2020) are computationally expensive and further lack a connection to counterfactuals. For a detailed discussion, see Hammoudeh & Lowd (2022).

Given a query sequence, influence functions require computing $g_{query}^T H^{-1} g_i$, where $H$ is the Hessian matrix, $g_{query}$ is the gradient of the measurement (query) function, and $g_i$ is the gradient of the training loss for a specific training example. There are two major computational challenges. Firstly, the formulation necessitates computing the Hessian matrix and its inverse. Since the Hessian has the same dimension as the model parameters, it is infeasible to explicitly compute and store it for LLMs with billions of parameters. Secondly, after computing the inverse-Hessian-Vector-Product (iHVP) $H^{-1} g_{query}$, the formulation requires calculating the dot product with all candidate training gradients $g_i$. This computation is equivalent to the cost of pre-training the model (slightly more expensive, as it requires computing the per-sample gradient). While Grosse et al. (2023) introduced a mechanism to efficiently approximate the Hessian, the primary computational bottleneck arises from computing the gradients of the candidate training sequences. This bottleneck poses a significant challenge for applying influence functions to LLMs.

One potential path to achieving more feasible computation is gradient compression (e.g., low-rank approximation). We can cache the compressed gradient of each training sequence on disk to avoid recomputing it. It is worth exploring the use of an approximate nearest neighbor search (Johnson et al., 2017) with these compressed gradients. Park et al. 2023 use random projection to compress the gradient, while Kwon et al. 2023 impose a low-rank structure to the gradient. As the gradients are now represented in a lower dimensional space, these approaches can also reduce the cost of performing IHVP. However, the above approach still requires computing the gradient for each training sequence. One way to avoid this is to save intermediate gradients during pre-training. This can potentially slow down the pre-training process but avoids the need for recomputation. A limitation of this approach is that the gradients are computed on the fly at the given parameters during training, not the final parameters, which deviates from the original influence functions derivation.

In summary,

- **Retraining-based methods:**
  - These methods would yield the most accurate attribution results, as

they directly approximate the counterfactual by retraining models with different subsets of the training data. However, they are not feasible for large language models (LLMs) due to the significant computational overhead of repeatedly retraining these massive models.

- Despite their limitations for LLMs, retraining-based methods can still serve as a valuable ground truth for smaller models and datasets, providing a benchmark for evaluating the accuracy of other attribution techniques in these settings.

- **Representation-based methods:**
  - These methods can be computationally and memory-efficient compared to other attribution techniques, as they only require forward passes through a neural network to obtain hidden representations.
  - However, their connection to the counterfactual notion of how the model's behavior would change if certain training data were removed is unclear. There is no obvious way to formulate this relationship, and further research is needed to determine if representation-based methods can serve as reliable tools for TDA.

- **Gradient-based methods:**
  - Among the existing gradient-based methods, influence functions are the most feasible option for TDA in LLMs.
  - However, directly applying influence functions to pre-training sequences is not feasible, as it requires computing the gradient for all pre-training sequences, incurring a computational cost equivalent to pre-training the model itself.
  - The most feasible approach is to save low-dimensional representations of the gradients on disk and perform a similarity search whenever a new query sequence is introduced. This technique leverages gradient compression and approximate nearest neighbor search to alleviate the computational burden.

## How plausible is implementing efficient, accurate TDA on frontier LLMs? How much work would it be?

As previously discussed, there are three types of approaches to TDA - retraining-based methods, representation-based methods, and gradient-based methods. It currently appears that gradient based methods (using influence functions) are the most likely approach to implement efficient, accurate TDA at scale.

As a result, this report will focus on evaluating the plausibility of influence functions as a solution for TDA. Recent research on influence functions suggests that there exists a path to implement AI provenance with plausible

## How plausible is implementing efficient, accurate TDA on frontier LLMs? How much work would it be?

time complexities on even the largest LLMs via compressed-gradient methods.

Let N be the number of training data sequences, and P be the number of parameters in a frontier LLM model. Currently, frontier AI models such as Llama3 have on the order of ~100 billion training examples, and ~100 billion parameters.

Without modern optimization techniques, gradient-based TDA methods would require reconstructing the entire set of gradients (calculating one gradient per model parameter), for all training examples, for each query output to be tested. In simple English, that suggests that calculating TDA for a single output is $O(N * P)$, which is unfathomably large.

With recent optimization techniques such as gradient compression, it is plausible to calculate a reduced set of ~16k gradients (as an example number) once per training example. Then, this set of ~16k gradients can be stored for each training example, taking up approximately $16k * O(N)$ storage costs. For example, if there are 60 billion pre-training sequences and the low-dimensional representation of the gradient is 16,384, approximately 1 petabyte (PB) of memory would be needed to store these gradients.

Therefore, the upfront time complexity of generating all compressed gradients for a model is roughly on the order of conducting a single epoch of training. The storage costs for such compressed gradients is significant, but is still plausible for large AI labs.

For each query output of an LLM, a similar set of ~16k gradients must be calculated once. Then, this set of gradients can be compared to the stored gradient database to find the most similar training examples, using optimized techniques such as approximate nearest neighbor search. We can estimate that such an optimized search would take roughly $O(\log(N))$ iterations.

Therefore, the inference time complexity for calculating TDA on a single query output may be around $O(\log(N))$. This is likely to be substantially slower than generating the query output itself, but still feasible to calculate efficiently by large distributed compute systems such as those used by leading AI labs.

**Using these numbers, it's reasonably clear that the time complexity and storage costs of TDA are not implausible to conduct on frontier AI models. Training costs are roughly on par with existing costs to pre-train frontier AI models. Storage costs may be on the order of single-digit petabytes, which is feasible for large AI labs. However, TDA inference may still be slower than traditional inference.**

It must be emphasized that while these techniques are theoretically plausible, they have not been applied in the real-world to frontier LLMs. There may be unpredictable issues when scaling such processes up.

Most importantly, using these optimization techniques results in necessary decreases in the accuracy of TDA results. At this time, it's still unclear how the accuracy of TDA is impacted by optimization on massive-scale LLMs - no such research has been conducted. Early results on small-scale LLMs are promising, demonstrating competitive accuracy (Keun Choe et. al, 2024, Park et. al, 2023).

Furthermore, scalable evaluation techniques for assessing the accuracy of TDA results after undergoing optimization do not yet exist. That is, even if TDA is conducted on such large models, there are not yet good methods to determine *how* accurate the results are compared to the "optimal" result. Potential counterfactual evaluation approaches include subset removal, creating a linear data modeling score, and proxy task evaluations (Ilyas et. al, 2022, Park et. al, 2023, Bae, 2024).

## What would the timing look like for achieving efficient, accurate TDA on frontier LLM models?

Before TDA techniques will start to be tested on large frontier LLMs in production, TDA researchers will need to demonstrate that the accuracy of such techniques remain high despite optimization techniques, and that evaluation techniques to prove such accuracy are both effective and scalable.

Of course, estimates of the progress of such AI research are quite fuzzy. Timing depends significantly on the results of each incremental research step, and the amount of funding and researchers allocated.

Currently, extrapolating from the approximate number of researchers in the field, increased interest in the machine learning community, and the rate of progress in research, we might expect that we will have strong additional evidence (beyond Keun Choe et. al, 2024) that TDA techniques maintain high levels of accuracy despite optimization in 1 - 2 years. The development of evaluation techniques such as subset removal, creating a linear data modeling score, and proxy task evaluations may take in parallel approximately 1 - 2 years.

After such fundamental research is conducted, there should be sufficient evidence that TDA is a reliable enough tool to attempt scaling up to frontier LLM models, similar to Anthropic's recent scale-up of mechanistic interpretability techniques. Such research would need to be conducted almost entirely within a leading AI lab, as the only organizations with access to frontier LLMs and resources for such AI researchers. Such research may take at least another year to complete, but could easily take 2-3 years depending on results and interest from AI labs. A successful result would involve the development of new research insights (see research benefits discussed below), and proof that TDA inference at frontier-LLM scale is good enough to be used as a research tool.

In summary, it would probably take a minimum of 2 years of additional research to demonstrate efficient & accurate TDA on frontier LLMs for internal research purposes. It could easily take up to 4 - 5 years, depending on factors described above.

Providing public API access to TDA inference, once effective TDA has been achieved, would be largely a strategic decision by AI labs bottlenecked by compute costs, compute availability, and commercial incentives. Technically, it would be extremely feasible and rapid. With motivation by AI labs, it could take no more than 3-6 months to go from provably effective TDA to public deployment.

There is some evidence that researchers in TDA are achieving successful results beyond what is currently well-known in the AI research space, and that AI researchers and lab leadership may be generally underestimating the feasibility of TDA at scale. Multiple independent AI researchers we spoke to were unaware of recent advancements in TDA optimization, and the topic has had significantly less coverage than its sister domain of mechanistic interpretability. It's possible that greater awareness of recent successes may rapidly accelerate investment and attention into TDA, shortening timelines from the dates mentioned above.

## Does conducting TDA on fine-tuned LLMs make TDA techniques more feasible?

Conducting TDA techniques on fine-tuned LLMs is substantially more feasible than on its base foundation model. Fine-tuned models (specialized models with custom data and training examples built on top of much-larger foundation models) have roughly similar steps to train, store parameters, and conduct inference as their foundation models. However, they are trained on orders of magnitude fewer training examples - perhaps millions or low billions, as opposed to ~100 billion.

Because the cost of TDA is primarily dependent on the number of training examples, this means that TDA is correspondingly orders of magnitude more efficient. This has the following effects on improving TDA with gradient-based methods:

- A significantly smaller set of training examples (many orders of magnitude less) means that computing and caching the gradients for each training example has substantially less compute and time cost.

- Because there is less to compute and store, gradients can be stored with less compression (e.g. 16k -> 64k gradients per training example), improving the accuracy of TDA.

- Because there are fewer training examples, computing the nearest neighbor query can be done more accurately - instead of approximate nearest neighbor search, developers may be able to run exact nearest neighbor search.

Beyond gradient-based methods, analyzing a small enough set of training examples may even make retraining-based TDA methods feasible. Retraining-based methods are computationally expensive but extremely accurate, providing a direct counterfactual (i.e. causative) relationship between training data examples and model outputs.

As a result, we should expect that TDA techniques will begin to be conducted for fine-tuned models much earlier than for foundation models, and that such applications will be more widespread and accessible. Discussions with TDA researchers suggest that such applications of TDA are plausible to be conducted today for AI research, though the accuracy of such results is still uncertain.

Furthermore, conducting TDA on the training data for a fine-tuned model dramatically simplifies the training data disclosure issues inherent in providing TDA for frontier LLMs. That is, it may be infeasible to convince AI labs to disclose their entire training data corpus (as we discuss later), but relatively more feasible to convince the developer of a fine-tuned model to disclose their fine-tuning data corpus.

## Assume that TDA can be conducted efficiently on large models with reasonably high accuracy. What are the benefits to AI researchers?

The ability to conduct TDA would have numerous, significant benefits for AI researchers in terms of building more accurate, efficient, and aligned models. In fact, the benefits are meaningful enough that AI labs appear well incentivized to independently conduct research on TDA, as is evidenced by recent work led by Anthropic researchers (Grosse et al., 2023).

Furthermore, using TDA tooling for internal AI research is significantly more feasible than using TDA for publicly-enabled inference, for the following reasons:

- The number of queries for internal AI research is likely significantly less than for production APIs, making it less expensive to run for AI labs.
- Internal AI research would not need additional permissions to access training datasets, which are closely guarded by AI labs.

The benefits of TDA on developing frontier LLMs would include, among others:

### Mitigating the prevalence of false claims and identifying contributing sources

- TDA would be massively beneficial for researchers attempting to mitigate hallucinations and incorrect information. Incorrect responses could be linked to the exact training examples most responsible for that response.

- Debugging and analysis for individual hallucinations could be significantly improved, and overall patterns could be identified.

### Identifying training data that produces poor results (issues such as bias, misinformation, toxicity), improving data filtering / selection

- By identifying the training data responsible for both good and bad responses, researchers would be able to tell which sets of data produce poorer results.

- As a result, researchers should be able to see significant improvements in choosing effective and useful data. They will be able to identify datasets that contribute better to desired goals moving forward, or choose datasets that look similar to other highly effective datasets.

- Such improved data filtering and selection will lead directly to **shrinking overall model size / improving efficiency**

### Shrinking overall model size / improving efficiency

- By improving their data filtering / selection capabilities, researchers should be able to see significant improvements to reduce the size of models while maintaining overall accuracy and capabilities. This will speed up research and development, inference speed, and also reduce model costs.

- Models built using TDA analysis will likely be smaller, more efficient, and have equal or more accuracy than direct competitors that aren't using TDA.

- Currently, optimizing models has taken a backseat to scaling compute and training data as rapidly as possible. As the field matures, it's likely that tools like TDA will become a competitive advantage for certain AI labs.

### Improved mechanistic interpretability & alignment

- Researchers will have a much stronger understanding conceptually of which sets of training data are most responsible for certain behaviors (e.g. deception, lying, dishonesty).

    - The ability to conduct and interpret TDA is likely one of the pillars of effective interpretability research, or understanding "why the LLM model does what it does".

    - The other pillar is <u>mechanistic interpretability</u>, or understanding how internal weights / parameters of LLMs contribute to responses. Unlike TDA, mechanistic interpretability has <u>received significant attention recently</u> as a potential tool to help solve the "alignment" problem.

- TDA could help providers better align LLMs with human values and preferences, by tailoring the data mix.

- TDA can provide useful answers to interpretability research.

    - For example, an outstanding research question for math-enabled LLMs is: to what degree are LLMs "memorizing" vs. "computing" answers? If

an LLM generates a correct answer, did it evaluate the question itself or simply see enough similar questions & answers online? TDA would provide the tools to answer this.

### Improved model customization and editing

- Recently, Anthropic demonstrated that they can emphasize certain attributes of a frontier LLM by increasing weights that correspond to specific concepts. Similarly, TDA will allow developers to emphasize desired attributes by weighting specific training examples more heavily.
- AI labs could use TDA to design custom versions of models that are more focused on certain topics (e.g. technical analyses) or have specific desired attributes (e.g. creative writing ability) by altering the training data set.

## What is the main bottleneck preventing public access to TDA inference for frontier LLMs?

On top of providing significant internal benefits to researchers at AI labs, it is plausible that TDA inference could be exposed as a tool for external parties to use - e.g. for corporations developing fine-tuned models, or even individual users calling APIs. Let's call this use case "public access to TDA inference".

Even if TDA inference becomes efficient and accurate enough, there is a significant bottleneck preventing public access to TDA inference for frontier LLMs: AI labs' unwillingness to expose their training datasets.

Currently, no leading AI labs produce frontier AI models with publicly available training datasets. Even Llama3 by Meta, an open-weights model, does not publicly share training data or sources of training data. AI labs may occasionally publish training datasets as part of a transparency or commercial initiative, such as Google Cloud's public datasets – however, these are not sufficient to train frontier AI models.

This is the case for several reasons:

1. **Competitive advantages:** The training datasets used to generate frontier LLMs is a key competitive advantage, as it constitutes part of their "secret sauce". It is well established that good training data curation substantially improves model quality (Zhou et al, 2023).
2. **Proprietary data:** In many cases, the training data includes proprietary or licensed information that the labs do not have the rights to distribute publicly. This could include data from books, websites, or other sources that are subject to copyright or terms of service restrictions.
3. **Privacy:** Training data may include sensitive personal data that could be used to identify individuals, such as logs from ChatGPT users. Such data cannot be shared publicly for legal and ethical reasons.

4. **Legal liabilities:** Releasing training datasets increases exposure to legal liabilities from data collection and usage. AI labs are actively defending their right to use publicly available copyrighted works in their training data. Public access to data sources would exponentially increase the number of opt-outs and legal challenges.

Without the incentives or ability to expose training datasets, AI labs will be similarly unwilling to expose public access to TDA inference for the corresponding models, as TDA inference returns specific training data examples.

There are several ways to bypass this issue:

- **Selective TDA inference:** AI labs may flag certain types of training data as available to be publicly shared (e.g. Common Crawl or Wikipedia datasets). They could selectively remove results from TDA inference when the training datasets are not publicly available.

- **Returning source metadata rather than the exact training data itself:** Some privacy and proprietary data concerns may be mitigated by simply returning the source of the data, such as the title of the book, research paper, or website, rather than the exact text itself.

- **Expose TDA inference for non-frontier AI models:** AI labs may selectively choose to release less powerful models with fully public training datasets, such as EleutherAI's recent models trained on the Pile.

- **Legal requirements for disclosure:** It is possible that governance policies may demand that AI labs publicly disclose their training datasets. This seems unlikely, for the competition and privacy reasons described above.

Of these options, our intuition is that a combination of **selective TDA inference** and **returning sources rather than exact training data** would be the most likely approach to bypass this bottleneck if AI labs were otherwise incentivized to provide public access to TDA inference.

## Assume that TDA is efficient, accurate, and AI labs willingly provide public access to TDA inference. What high-level benefits might we expect?

### Preventing copyrighted data usage

- If AI labs provided public access to TDA inference on their entire training dataset (rather than selectively permitting data), it would become significantly easier for content creators to identify copyright violations regarding their work.

- Though societally beneficial, this is a serious disincentive for AI labs to enable public access.

### Fact checking / content moderation

- TDA could improve public fact-checking of responses by tracing claims back to original data sources for verification (Akyürek et. al, 2022). This would provide protection against hallucinations and false claims.
- Publicly available tools could be created for users to access TDA inference.
- It would enable better accountability for AI models to produce accurate responses, and a method to "double-check" the LLM.

### Trustworthiness and user confidence in LLMs

- Whether or not this would improve user trust in LLMs is unclear, and likely depends on the model / implementation.
  - Users may gain trust because they can verify answers themselves, or because they have more transparency into the rationale behind LLMs.
  - They may also lose trust when they must invest time to confirm their LLM responses, or when they find their responses are incorrect.
- Even if user trust may be debatably impacted, the overall *trustworthiness* of models should strictly improve due to TDA techniques.

### Accelerating external research

- The same research benefits as described previously would be available to external researchers to evaluate, democratizing and accelerating benefits from TDA.

### Providing accountability for AI labs

- AI labs would have greater external accountability for:
  - The quality, diversity, and accuracy of their LLM responses
  - Responsible training data collection and curation
  - Protecting intellectual property rights and paying creators fairly

---

## Are AI labs incentivized by these benefits to provide public access to TDA inference? Would a government choose to mandate access?

As mentioned in the previous section, there are many societal benefits to exposing public access to TDA inference. However, many of these are related to transparency, accountability, or conducting external research, rather than providing practical functionality to individual users. It's not clear that there exists meaningful demand for such a tool by people not working directly on AI research or development.

Furthermore, many of these outcomes are societally beneficial, but disadvantageous for AI labs. For instance, TDA inference enables holding the AI lab more accountable, which is a clear disincentive to expose such a feature. As a result, it's not clear that AI labs have aligned incentives to expose public access to TDA inference - instead, they seem reasonably strongly disincentivized.

In the case AI labs choose not to expose their training datasets or TDA inference, it seems unlikely that regulators will step in to mandate public access to either.

- Exposing training datasets has many practical privacy and competitive issues as mentioned, and governments would be reluctant to get directly involved in AI competition.
    - For example, the US government is highly disincentivized to force US companies to expose their training data publicly, because that would significantly benefit AI research for Chinese competitors.
    - The EU may approach this problem somewhat differently, as it's more focused on protecting citizen's rights. There are some public benefits to exposing training datasets such as transparency and accountability. However, any attempt to force AI labs to release training data would face significant legal and industry pushback.
- TDA inference is both niche and research-focused, which is well beyond the scope of regulators. They have no obvious incentives and little upside to require that AI labs mandate public access to TDA inference.

**It seems most plausible to me that a few AI labs focused on research and transparency might expose such a feature publicly as a competitive advantage, but that the majority of AI labs would primarily choose to use TDA inference privately to improve their models.**

## What are policies and systems enabled by TDA inference?

We'll describe a variety of plausible policies & systems enabled by TDA inference. This list is not exhaustive, but covers many of the most likely potential impacts.

Because some of these projects are the combination of a policy requirement combined with a system, we've grouped all of these into the same category. We'll note the type of each project as "Policy & System", "System", or "Policy".

As discussed before, public access to TDA inference is a significant outstanding question. We'll note for each project whether or not it requires public access or auditor access to TDA inference.

## Royalties to Data Providers / Creators

**Type:** Policy & System

**Note:** Would not require public access to TDA inference, as this system could be deployed and conducted internally within an AI lab. Auditor access may be required to verify compliance with any regulations.

Presumably, effective TDA would create a clear relationship between LLM responses and the training examples most responsible for that output. For example, a generated image for the prompt "create a Cubist painting of X" might show that it drew influence primarily from paintings created by Picasso.

Given TDA techniques that could be run efficiently at scale, it is both plausible and feasible to design a system that could give data providers credit for their influence on the outputs of LLMs. Should sources of training data for LLMs receive further protections in terms of permitted use, it may be to the benefit of both AI labs and data providers to develop a royalties system.

Such a system could provide micropayments to creators in return for the inclusion of their data into an LLM, in the same way that Spotify pays royalties to artists for inclusion of their music into their product. Micropayments could correspond to the amount of influence creators had on a LLM response (Keun Choe et. al, 2024). However, the social impacts of such a policy would need to be evaluated.

This system would offer an alternative or supplement to upfront AI content licensing deals, such as Google's deal with Reddit or OpenAI's contract with Dotdash Meredith. Rather than paying for content upfront, payments could be deferred and made based on real-world usage by LLM models.

Currently, training generative AI models on copyrighted works available online is considered fair use, and AI labs are defending their right to use copyrighted works in their training data. Whether this will continue to be the case is a matter for the courts.

**If there is a landmark judicial ruling declaring that AI labs were not permitted to use copyrighted works available online without the consent of the creator, such a royalties system would probably become highly attractive to AI labs.** It would allow labs to incentivize creators to permit LLMs to train on their data, while avoiding negotiating or paying upfront AI content licensing fees.

Without such a ruling, it is still possible that AI labs may see positive incentives to develop such a system, either for positive public feedback, to avoid lawsuits, or out of an ethical responsibility.

## Automated Response Improvement / Fact Checking

**Type:** System

**Note:** Would not require public access to TDA inference, as this system could be deployed and conducted internally within an AI lab.

If TDA became efficient enough to use in the inference response pipeline for LLM providers, it's quite likely that it will be used as an automated supplementary tool to enhance and improve LLM responses. This would not require public access to TDA inference.

Such a system would work similar to the following description:

- For any LLM response, the system would conduct TDA inference and identify the 3-5 most influential sources of training data.
- These sources of training data would be used in an automated fashion to do the following:
    1. Potentially improve the quality of the LLM response by surfacing and integrating additional information missed by the first inference step
    2. Identify if the LLM response is obviously wrong or misaligned with factual data. Or, it could identify if the primary source of the response is unreliable (e.g. Reddit) and the results should be discounted.
- The results from this TDA analysis would be combined with the original LLM response to produce a potentially better LLM response, and served to the user.

If able to be conducted efficiently and with meaningful improvements, such a system could serve as a competitive advantage for AI labs developing mature LLM systems. As a result, this is likely to happen once TDA is both efficient and accurate enough to be useful.

### Tooling for Improving External Audits of Training Data

**Type:** Policy & System

**Note:** Would not require public access to TDA inference. However, it would require auditor access, which would be an external party to the AI lab.

It is plausible that future regulations may require that AI labs disclose their entire training datasets, either publicly or privately to a third-party auditor, to confirm they are compliant (e.g. not in violation of copyright law, non-discriminatory, etc.). If this was enforced by third-party auditors tasked to evaluate the applicable training datasets, TDA inference would be a useful tool.

Specifically, governments could mandate that qualified external auditors be granted access to run TDA queries on AI models in order to examine the training data. Instead of needing to evaluate the entire dataset, this would allow auditors to test LLM responses to identify specific data sequences.

As with most auditing practices, challenges would include protecting legitimate trade secrets and ensuring secure access for auditors.

### Could TDA techniques be used as an alternative for RAG for content attribution?

**Type:** System

**Note:** This would require public access to TDA inference.

Currently, retrieval augmented generation (RAG) is the primary tool used by AI models to enhance LLM prompts with relevant data to provide accurate, practical responses incorporating information sources external to the model. In short, RAG-enabled LLMs search a vector database for relevant results, augment those results to include additional info, and generate LLM responses based on the final results. Importantly, RAG-enabled LLMs can reliably handle "content attribution": <u>citing and linking the source of their external data</u> to provide additional context and fact-checking.

Presumably, TDA techniques may provide a substantially different, but potentially powerful approach to handle "content attribution" for information generated by an LLM response. Rather than linking to an external document as in RAG, TDA may allow identification of specific training examples that most impacted the LLM results. A citation or link to the source of the training example could be provided.

However, TDA techniques have many limitations compared to RAG as a method of source retrieval for LLMs.

1. **Accuracy & Precision:** TDA measures the influence of training examples on the model's outputs, but this influence can be complex and indirect. A model's output may be influenced by multiple training examples in combination, rather than a single direct source.
   a. As a result, it may not consistently return relevant sources, or may distribute influence across too many sources to be useful.
2. **Speed:** TDA would likely be slower than RAG, as performing nearest-neighbor search would take longer than searching an optimized vector database.
3. **Granularity:** TDA typically attributes influence to entire training examples or passages, rather than pinpointing the exact sentence or phrase that contains the desired information. If the relevant information is just a small part of a larger training example, TDA may not provide the necessary granularity to identify the specific source.
4. **Maintainability:** RAG is easier to keep up to date, as it is much easier to update a vector database rather than continually retraining a new model.

Most importantly, TDA cannot be directly compared to RAG as they have fundamentally different methods of retrieving data. The responses of TDA are tightly coupled with LLM training data, whereas RAG is tightly coupled with domain-specific vector databases.

From evaluating these two technologies, we believe **content attribution tooling for RAG and TDA would occupy parallel and different use-cases.**

TDA could be used to link users of general-purpose LLMs to citations as a general-purpose tool for discovery or fact-checking answers. These citations would not always be logically understandable from a human perspective (as they rely on the inner workings of LLMs), and may be slower.

Meanwhile, RAG would be more effective at retrieving specific information from a purpose-built, targeted database. If effective content attribution was a key use case of an LLM, it's more likely RAG would be implemented.

## Mandatory Public Access to TDA

**Type:** Policy
**Note:** This would require public access to TDA inference.

Regulations could require that AI companies provide users with access to the top attribution results from TDA for their queries, in the interest of transparency.

As discussed before, this is unlikely. AI companies would strongly resist this as it could expose details about their proprietary training data and models, and would require significant additional costs. Regulators are typically uninterested in such a niche tool, and this would be considered a significant overreach of regulatory power.

---

# How could TDA contribute to mitigating large-scale risks from AI systems?

**Disclosure:** The following research was produced by researchers from Convergence Analysis, originally on behalf of the Future of Life Foundation. Both organizations have core mission statements centered on the mitigation of existential and large-scale risks from AI systems.

Advanced AI systems may pose significant large-scale and extreme risks to society. Examples of these may include societal risks such as AI-triggered political chaos and epistemic collapse, physical risks such as AI-enabled biological, nuclear, or cyber catastrophic risks, and existential risks such as loss of control of or to future superhuman AI systems[3].

In this section, we'll briefly evaluate TDA's potential effect on such large-scale AI risks, should widespread adoption of TDA come to pass. For this analysis we'll consider three different domains of impact: societal impacts, AI capabilities, and technical AI safety.

## Impacts on Society

There are a number of direct societal and systemic effects from widespread adoption of TDA. We'll discuss the importance of a few effects below:

- TDA may help reduce hallucinations and serve as a useful supporting measure for reducing false information from LLMs.
  - *Epistemic security* (access to reliable information) is a potential risk factor from AI systems. However, threats to epistemic security are currently driven by a wide range of social and systematic sources (e.g. recommendation algorithms, clickbait & outrage content, polarized news sources). Reducing LLM hallucinations will reduce an emerging new source of false information, but likely does not improve these existing sources.
- TDA research may somewhat improve the trustworthiness of LLM models and increase public accountability, e.g. via third-party audits of training datasets.

---

[3] The description of these risks was borrowed directly from Our Position on AI - Future of Life Institute.

- Audits and accountability could improve societal outcomes from LLMs, by reducing issues such as discrimination or copyright violations.
- Increasing trustworthiness has various positive societal benefits, such as more widespread use of AI in beneficial applications. On the other hand, it might lead to overreliance on AI systems.
- Protecting intellectual property rights and paying data creators fairly is a key potential benefit.
  - This could possibly form a small part of a positive solution to the massive expected economic impacts from TAI.

Overall, we believe that improved TDA may have some positive societal impacts, and may be a contributing factor to reducing some societal sources of large-scale risks. However, we must note that TDA doesn't seem to be a critical factor for any specific source of risk.

## Impacts on AI Capabilities

Many of the most tangible effects of TDA appear to be largely related to AI capabilities and research - benefits such as identifying the source of poor LLM responses, reducing hallucinations, and shrinking model sizes.

In the short-to-medium term, this type of research may actually improve AI capabilities by allowing models to be smaller and more efficient, decreasing the number of training examples required to produce a model with similar capabilities.

As a result, AI labs likely have strong incentives to fund TDA in the near future because of the potential benefits to capabilities research and alignment research. Currently, it doesn't seem to be a priority for AI labs, in large part because the techniques have not yet demonstrated feasibility at scale. However, this could change rapidly in the next 2-3 years as more research and results come out.

Because increasing AI capabilities are directly associated with increasing risks from AI systems, efforts to improve TDA may actually somewhat increase large-scale risks from AI systems in the short-to-medium term by accelerating AI development.

## Impacts on Technical AI Safety

It's quite possible TDA could be an important component of technical AI safety - specifically, for understanding the impact of training data on the alignment of AI models. It has many parallels to existing mechanistic interpretability research, which is largely focused on understanding the impact of parameters on model outputs. In fact, TDA should be considered a sister branch of mechanistic interpretability research, focused on training data rather than weights.

For example, TDA may help us answer questions such as:

- Is this model referring strongly to training data examples that display

concepts of "deceptiveness" or "power-seeking"?

- Are certain types of queries more likely to lead to training data retrieval from low-value sources (such as Reddit)?

We believe that TDA is a promising form of technical AI safety research, and that it may make meaningful improvements to AI alignment - perhaps comparable to the impact of *mechanistic interpretability* research. As a result, we think that efforts to improve TDA may have long-term positive effects, because of its potential to decrease large-scale risks from AI misalignment.

### Summary of Impacts

We expect the following outcomes regarding TDA's impact on mitigating large-scale risks from AI systems:

- Marginal benefits in terms of TDA's societal impacts
- Clear negatives due to TDA's likelihood to improve AI capabilities
- Substantial potential benefits due to TDA's likelihood to improve AI alignment

In short, there isn't a conclusive answer, but rather we believe that TDA's impact on large-scale risks depends on one's relative weighting of different priorities. In the short to medium term, accelerating AI capabilities may increase risks from AI systems. In the long run, we believe that TDA's benefits to AI alignment will likely outweigh its drawbacks in accelerating capabilities.

We also believe that there is a moderate likelihood that this type of research will be soon funded by AI labs interested in capabilities and alignment - Anthropic, in particular, seems to be a likely candidate due to their current mechanistic interpretability research.

# Conclusion

Research on training data attribution (TDA) appears to be progressing steadily towards becoming efficient and accurate enough to be practical. Promising results show that techniques exist to substantially reduce compute costs while maintaining accuracy. However, these techniques have not yet been applied to larger models. Better evaluation metrics also need to be developed before TDA inference will begin to be useful.

Looking ahead, we expect to see increasing investment in TDA research from leading AI labs, driven primarily by its potential to improve model development and alignment research. The technology's ability to enhance model efficiency and provide insights into model behavior makes it a compelling area for investment, despite current implementation challenges. While public access to TDA inference may remain limited due to data privacy and competitive concerns, the technology is likely to become an important internal tool in the AI research ecosystem.

The next 2-5 years will be crucial in determining whether TDA can fulfill its promise as both a practical tool for AI development and a contributor to AI safety research. Initially, we'll likely see TDA inference used to improve smaller, fine-tuned models. As research and techniques improve, we'll likely start to see TDA inference used on large / frontier LLM models. In both cases, TDA will provide meaningful in-house research benefits.

Overall, one of the most promising aspects of TDA is likely its potential to improve the technical safety and interpretability of AI models. The ability to trace and understand the relationship between training data and model outputs appears to be a meaningful lever to developing safer and more reliable AI systems. As AI capabilities continue to advance, TDA may become increasingly important for ensuring that AI systems remain interpretable, accountable, and aligned with human values.